# A New Approach to Quantum Computing Multi-Qubit Generation and Development of Quantum Computing Platform with Magnetic Resonance Imaging Techniques

**Zang-Hee Cho[1*], Young-Don Son[2], Hyejin Jeong[1], Young-Bo Kim[3], Sun Ha Paek[4], Dae-Hwan Suk[5], Haigun Lee[5,6*]**

[1]*Neuroscience Convergence Center, Institute of Green Manufacturing Technology, Korea University, Seoul, South Korea*
[2]*Department of Biomedical Engineering, Gachon University, Incheon, South Korea*
[3]*Department of Neurosurgery, Medical School, Gachon University, Incheon, South Korea*
[4]*Department of Neurosurgery, Seoul National University Hospital, Seoul National University, South Korea*
[5]*Institute of Green Manufacturing Technology, Korea University, Seoul, South Korea*
[6]*Department of Material Science and Engineering, Korea University, Seoul, South Korea*
\* Correspondence: haigunlee@korea.ac.kr (H.L.); zhcho36@gmail.com (Z.-H.C.)

**ABSTRACT**

Increased interest in quantum computing has resulted in various proposals for generating quantum bits or qubits. Superconducting qubits of the Josephson junction are the most widely accepted and currently used, while ion-trap and similar molecule-based qubits have been proposed more recently. In these methods, each qubit is generated individually with great effort. Here, we proposed a new technique using magnetic resonance imaging (MRI) to simultaneously generate multiple qubits and provide a complete qubit platform for quantum computing. Central to this method is the "gradient" concept with multiple radiofrequency coils, one for all qubits and others for individual qubits with each small Q-coil. Another key concept is the time-encoded probability amplitude technique, using individual Q-coils with the spin-echo series in each qubit incorporating readout gating for time-encoding. Our newly proposed MRI-based qubits are well-suited to currently available electronics, superconducting and semiconductor technologies, and nuclear magnetic resonance and MRI physics and technology.

## 1. INTRODUCTION

One of the fastest growing subjects in science and technology in recent years is "quantum computing," which is contrasted with the modern super-computers[1-12]. This new computing concept appears to be entirely different from the existing computing system: instead of "bits" used in classical computing, the basic computing unit in quantum computing are quantum bits, or "qubits." Qubits have two bases, that is, |0> and |1> exist simultaneously, with attached probability amplitudes, $\alpha$ and $\beta$. Interestingly, nuclear magnetic resonance (NMR) shares the same principles, making it suitable for quantum computing. In NMR or magnetic resonance imaging (MRI), the spin system has up and down spin populations, which represent the low and high energy states existing simultaneously[13-15].

MRI-based qubits (MR-qubit) follow quantum physics, which is a deterministic process but has a quantum statistical nature, such as the probabilistic distribution of the spins (up and down states). The key features of the qubit, in contrast to the currently used classical bit, which is deterministically based on "0" or "1," can be written as:

$$|\Psi> = \alpha|0> + \beta|1> \quad \ldots \quad \text{Eq. 1}$$

where $\alpha$ and $\beta$ are the probability amplitudes of the qubit state "|0>" and "|1>" or the upward and downward vectors, respectively, with continuously varying values of $\alpha$ and $\beta$. This statistical nature of the state vectors and varying probability amplitudes, $\alpha$ and $\beta$, has quantum mechanical properties, as found in many atomic-scale objects, such as isolated atoms and molecules. Most of the current qubits are based on this atomic scale property, for example, the superconducting Josephson junction or trapped-ion qubits[4,9,10].

A technique that is somewhat different, although it has the same principles, is the NMR-based technique, which utilizes nuclear spin distributions under a strong magnetic field[2,13-15]. NMR techniques are actively used in many areas of scientific and technical fields, such as NMR spectroscopy and NMR-based medical imaging, i.e., MRI, which has been developed in the last few decades and is widely used worldwide[16-18]. MRI techniques of various types are used in medical imaging, with highly sophisticated applications ranging from extremely fast imaging to diffusion and perfusion imaging that employ various pulse sequences with gradients and radiofrequency (RF) coils[19-25]. It should also be noted that the various high-field MRI scanners that have been developed in the last few decades, ranging from 1.5 T to 7 T or even 11.7 T, together with the associated MRI signal processing techniques of various types, ranging from the ultra-fast multi-spin-echo sequence to ultra-high-resolution fiber tractography that allows in vivo human imaging with a resolution down to 100 μm, appear ideally suited for quantum computation and its appliations[26-28]. Along the same line, MRI microscopy of small samples, down to the level of 1–4 μm, has also demonstrated the possibility of using MRI for future quantum computing applications[29,30].

*E-mail: zhcho36@gmail.com

Although ideas of utilizing NMR concepts are not new, it has not been pursued actively because of the overwhelming dominance of other techniques, such as the superconducting qubits and ion-trap techniques[4,10]. In this paper, we propose a new avenue for the qubit generation using the NMR concept and particularly as related to MRI techniques. NMR and MRI are certainly macroscopic approaches; therefore, the signals follow the rules of ensemble averages, rather than single atomic level behavior, for the generation of MR-qubits[7-9]. As will be described below, the proposed MR-qubit generation and its application have a few unique advantages and potentials for future quantum computing applications, namely:

1. "MRI gradient"-based multiple qubit (MR-qubit) generation is new, particularly when using different RF coils, namely one main RF coils for excitation of all qubits simultaneously, and activating selected numbers of qubits by using multiple small qubit coils (Q-Coils) for spin-echo generation and time-encoded echo-signal readouts. Interestingly, all underlying techniques are readily available and well-established and are currently in active use, particularly in high-field (1.5–7.0 T) clinical human MRI; therefore, it is a well-proven technology.

2. The readily available qubit state-generation technique, "time-encoded probability amplitudes (TEPA)," for each MR-qubit is a new concept. In this technique, it is easy to control and manipulate multiple well-encoded qubits and qubit states via the spin-echo technique with readout operation.

3. Although a strong superconducting magnet is required for MR-qubit generation, the MR-qubit sample itself (water or carbon) can be placed in a room temperature environment (H or $C^{13}$ bar), which allows easy control and manipulation if necessary. It should also be noted that, although the superconducting magnet requires a low temperature (4 K), the qubit itself does not require ultra-low temperatures, as does the superconducting qubit (which requires temperatures of 10–40 mK). MR-qubit generation also does not require an ultra-high vacuum environment, as does the ion-trap qubit system.

Among the advantages of MR-qubits for quantum computation, multiple qubit-generation with the MRI gradient technique and RF coils of various types, which are currently readily available, are highly advantageous elements. Moreover, the TEPA technique is another useful tool and advantageous component from a methodological view-point, as compared to existing techniques, such as superconducting qubits and ion-trap qubits, for future general purpose quantum computer appliations[8,9].

**NMR and MRI techniques for qubit generation**

One of the powerful and unique aspects of MR-qubit generation and its application to quantum computing is the easy availability of the related technology and its ready implementation with the presently developed quantum computing devices, such as quantum logic gates and algorithms[1,5,8,9]. Before commencing a description of qubit generation, we show standard MRI in a conceptual form in **Fig. 1**, for reference. We have shown both the main magnet and three gradient coils ($G_x(x)$, $G_y(y)$, $G_z(z)$) for imaging with an RF coil.

In **Fig. 2,** a qubit-generating platform similar to the clinical MRI system is shown**.** In the middle, an H-bar (proton or carbon bar, such as water or $C^{13}$, respectively) is placed under a high static magnetic field, $B_0$ (brown superconducting magnet). Under the strong magnetic field $B_0$ and gradient $G_z(z)$, with the main RF coil (large green coil) and a set of multiple small RF coils (Q-coils), a large number of qubits can be generated simultaneously, each of which can be controlled separately with the TEPA technique. Each qubit has its own RF coil (Q-coil), with its own frequency and its own finite frequency bandwidth. The gradient $G_z(z)$ and main magnetic field $B_0$ have the same direction.

Initially, all the qubits are under the main static or system magnetic field $B_0$, which forms the System Hamiltonian and a gradient $G_z(z)$. Unlike conventional MRI in quantum qubit application, only $G_z(z)$ is used. First, the initial control starts from excitation of the sample (all qubits) by the main control RF coil at $t = t_{0+}$; however, each qubit is separately controlled by the individually positioned Q-coils with different RF frequencies ($\omega_1$, $\omega_2$, etc.) and operated by a multiple spin-echo sequence for multiple spin-echo signal generation. Each echo signal has their own α and $β$ values, the probability amplitudes.

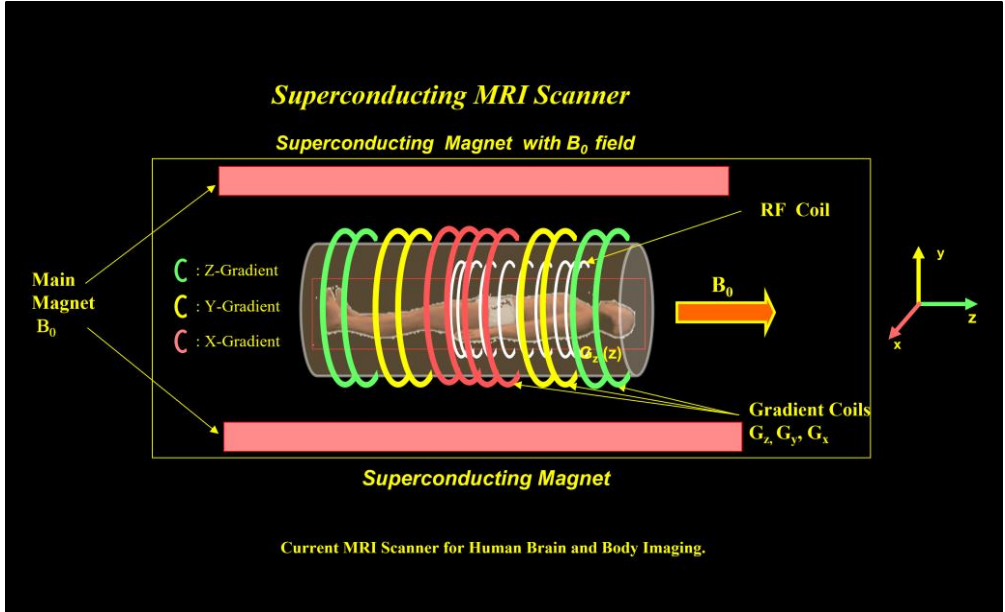

**Fig. 1 Current MRI scanner for human brain and body imaging.** Human magnetic resonance imaging (MRI) system with a main magnet with a field-strength of $B_0$, three gradient coils (Gx, Gy, Gz) with a radio frequency coil (RF coil) for excitation and signal readout. Note the direction of the main magnetic field $B_0$ and the direction of the subject (human body), which we denoted as the z-direction.

These spin-echo signals are read out by the readout gates (RG1, RG2, etc., are the readout gates for qubit 1, qubit 2, etc.) at a selected time ($t_{1+}$, $t_{2+}$, etc.). By properly chosen control time or selection time of the echo signal, one can control the TEPA, such as α1 and β1 for any selected qubit.

In **Fig. 3**, the spin-echo signal generated in a selected qubit is shown in relation to their probability amplitudes $α_s$ and $β_s$, which represent the qubit states where $α_i$ and $β_i$ are the probability amplitudes that are selected by spin-echo time $t_{i+}$ for each qubit. We assumed that the main RF coil is activated at time $t = t_{0+}$ and that any desired state can be selected for each qubit by the signal control time (selected spin-echo time or gate readout time), that is, $t = t_{i+}$, such as $t = t_{1+}$, $t = t_{2+}$, etc. We named this the TEPA method and used it for probability amplitude encoding by control time or spin-echo time for each qubit.

In actual practice, all the RF coils (the main RF and Q-coils), readout quantum gates, logics, and gradient are integrated on a microchip. In Fig. 3, T1 relaxation time (spin lattice relaxation time) and the possible spin-echo signals that will be generated during the relaxation time (which could be approximately 5 × T1) are shown. This coherence time, or the computationally useful time, is based on the simplified NMR or MRI spin-density signal that is related to the T1 and T2 relaxation times, as shown below:

$$\rho(x,y,z) = \rho_0(x,y,z)\left\{exp\left(\frac{-t}{T_2(x,y,z)}\right)\right\}\left\{1 - exp\left(\frac{-t}{T_1(x,y,z)}\right)\right\}$$

… Eq. 2

where T1 and T2 are the spin–lattice and spin–spin relaxation times of the sample (H-bar), and $\rho_0$ (x, y, z) is the initial spin-density value. The last term, the spin–lattice relaxation time or recovery time, which is also related to the spin-echo signal decay, is also shown in Fig. 3[13-15,18].

This spin–lattice relaxation or recovery time is an important component in quantum computational time, that is, the effective time for computation as well as generation of qubits. For example, since the T1 value of water at room temperature under a static magnetic field of 3 T is approximately 3500–4000 ms, the actual computation time could be as large as 20 s when we set the computation time as 5 × T1. During this time, the number of spin-echoes that can be generated by each qubit could be as large as 1000, provided that each spin-echo time interval, TE, is in the range of 20 ms. T2 is the spin–spin relaxation or dispersion time, which is usually smaller than T1 values (approximately half of T1) and affects the spin dispersion and refocusing of each spin-echo signal.

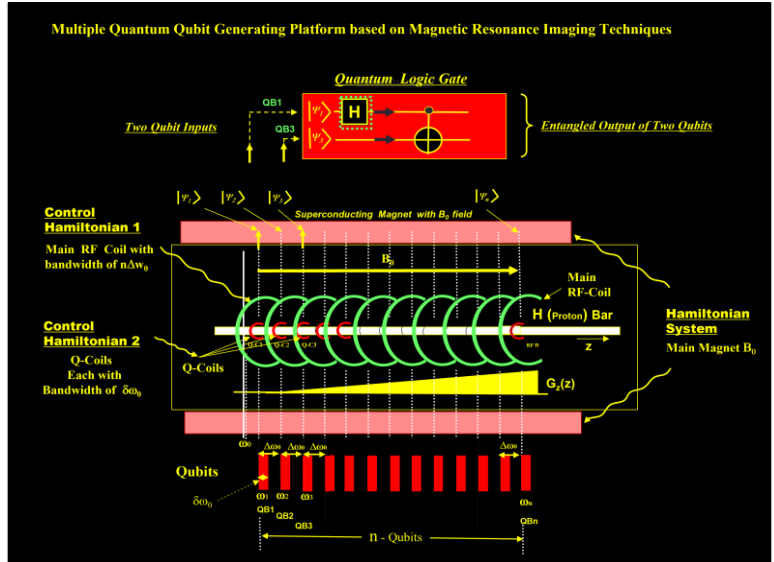

**Fig. 2 Magnetic resonance imaging (MRI) technique-based multiple quantum qubit-generating platform using an MRI gradient and multiple radiofrequency (RF) coils.** In the middle, a single H-bar (proton bar, such as water) is placed as a qubit-generating source. After the static magnetic field ($B_0$) is applied and system magnetization is established (System Hamiltonian), the next step is to excite all the qubits simultaneously by the main Control RF coil at $t = t_{0+}$ (Control Hamiltonian 1). Subsequently, each qubit (with small red coils) is separately controlled by the individually positioned Q-coils with different RF frequencies, $\omega_1$, $\omega_2$, etc. The desired qubit is excited by the corresponding Q-coil and each qubit generates a series of spin-echoes. Among these spin-echoes, according to the desired qubit state, one of the spin-echoes is selected, for example, at $t = t_{1+}$, $t = t_{2+}$, etc., and signal is by gating. Legends: $|\Psi_1\rangle$ and $|\Psi_2\rangle$ are the generated qubits, qubit 1 and 2, respectively. Q-coil: qubit RF coil. $\delta\omega_0$: one qubit RF bandwidth. $\Delta\omega_0$: qubit center frequency separation. $\omega_1$, $\omega_2$, and $\omega_3$: center frequencies of qubit 1, qubit 2, and qubit 3, respectively. $G_z(z)$: gradient in Z-direction. H: Water bar (proton bar). $B_0$: Main magnetic field. Q-C1, Q-C2, and Q-C3: RF coils or Q-coils for qubit 1, qubit 2, and qubit 3, etc.

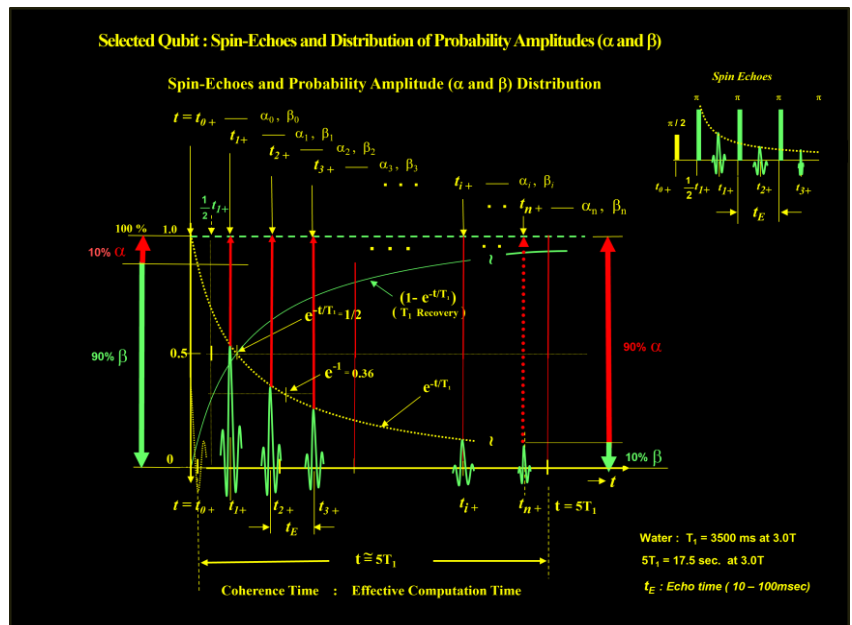

**Fig. 3 Spin-echoes and probability amplitudes (α and β) distribution.** Magnetic resonance imaging (MRI)-based qubit generation during the coherence or T1 spin lattice recovery time, which we assumed as the effective computation time. That is, during this time, the desired quantum qubit can be generated and computation can be performed. In MRI, this is equivalent to the T1 relaxation period, which we assumed to be approximately 5 × T1 ($5T_1$), as shown in the figure with the green curve ($1 - e^{-t/T_1}$) or the red dotted line ($e^{-t/\tau}$). During this time, MR

spin-echo signals can be observed; for example, multiple spin-echoes are generated in each qubit by the application of a series of 180° transverse RF pulses. The green spin-echo signals show decreasing trends of spin echoes representing the β portion of the probability amplitudes, while the α portion is increasing (red lines). From this, any desired qubit state, that is, depending on the desired values of α and β, can be determined by reading the spin-echo signals at the corresponding times, for example, α1 and β1 for $t = t_{1+}$ and α2 and β2 for $t = t_{2+}$. The qubit states are read out via the readout gates (e.g., RG1 for qubit1). This time-dependent variation of α and β in each qubit, i.e., the qubit state as a function of time can be exploited for qubit generation. This generation of time-dependent qubit states or probability amplitudes (α and β) of a qubit is utilized for quantum computation.

In **Fig. 3**, we also illustrate the spin-echo signals that are generated in each qubit and $α_i$ and $β_i$ as a function of time $t_{i+}$ is shown. The time between the initial main RF excitation time, $t = t_{0+}$, to the individual qubit control signal, the pick-up times for the readout gate (RG1, readout gate of qubit1, shown on the upper part of Fig. 3, as the green bar), such as $t = t_{1+}$, $t = t_{2+}$ etc., are the encoding time of each qubit state.

In **Fig. 4,** a representative qubit, the j[th] qubit, with spin-echo signal selected for a time at $t = t_{3+}$ is shown, that is, the spin-echo signal selected at the time $t = t_{3+}$ by opening of the readout gate, RG1. With this example, we have also shown other possible qubit states that can be generated, such as:

$|Ψ_1⟩ = α_1|0⟩ + β_1|1⟩$ and $|Ψ_2⟩ = α_2|0⟩ + β_2|1⟩$

As shown, depending on the selection of spin-echo time or selection time, $t_{i+}$, any set of desired probability amplitudes can be selected for a qubit. As mentioned earlier, this TEPA method is one of the key components of the proposed qubit generation for quantum computing. The TEPA technique can be used with a table that can either be generated or calculated for each qubit, for selection of the desired values of the probability amplitudes, as shown on the right side of **Fig. 4**.

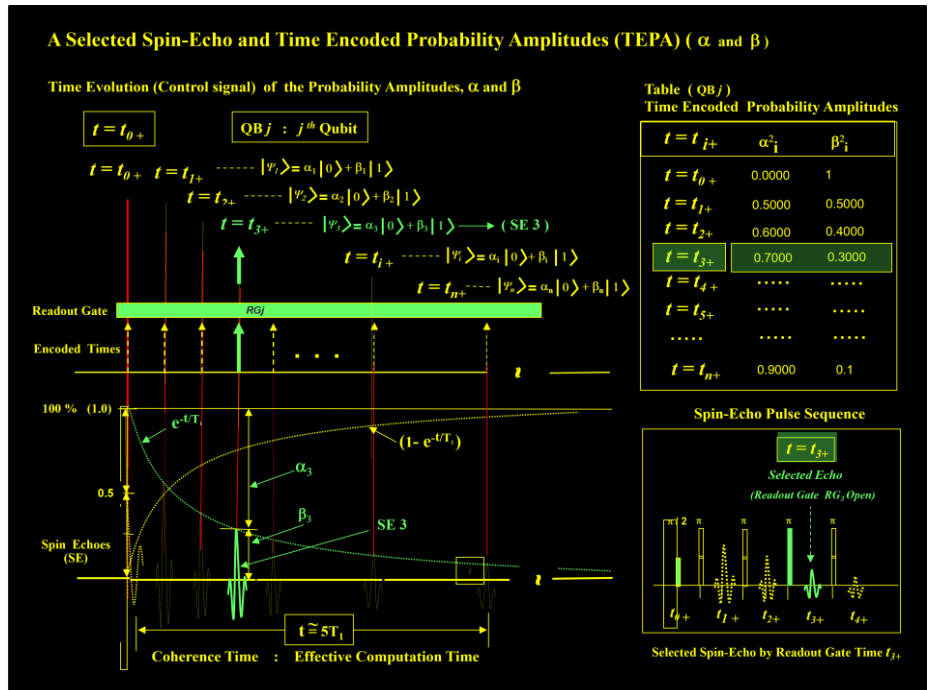

**Fig. 4 An exemplary illustration of a set of the selected control time-encoded probability amplitude (TEPA) values, α and β.** Probability amplitudes α and β are determined by the selection of the readout time, $t_{i+}$, for the readout gate (e.g., RG1 for qubit 1). In this example, RG1 is opened or read out at time $t = t_{3+}$, which has probability amplitude values of $α_3$ and $β_3$. Note here that the first excitation of the main control RF (90°) is the same for all the qubits, while the 180° spin-echo pulse series are only applicable to those qubits of interest for the generation of a series of spin-echoes. Based on the table on the right, one can select the desired probability amplitude values by selection of the gate readout times ($t_{i+}$). These TEPA values are the key component for the qubit state generation and is the tool for the quantum computation uniquely available from the proposed MRI-

based qubit generation. $|\Psi_1\rangle = \alpha_1|0\rangle + \beta_1|1\rangle$ and $|\Psi_2\rangle = \alpha_2|0\rangle + \beta_2|1\rangle$, etc., are the qubit states of qubit 1 and qubit 2, etc. At the bottom right, a typical spin-echo pulse sequence is shown for reference, in relation to the main figure on the left. In the top right, a table for the probability amplitude vs. spin-echo time or gate readout time is shown for reference. This table can either be constructed from calculations based on the T1 value, or from experimentally measured data in given environments. Here, also note the readout gate (green bar shown on the top left), i.e., RG1, which is the readout gate of qubit 1. SE3 is the spin-echo signal at qubit 3 for the readout time $t_{3+}$.

For a quantum computing application of MRI-based qubit generation and platform development, the desired number of qubits with appropriate probability amplitude values can be produced by suitable selection of qubits and the corresponding spin-echo time for each qubit. Selected qubits are then used for quantum computing with the quantum logic gates (see Fig. 2 as example). This is an exemplary quantum entanglement arranged in quantum computing with two MR-qubits generated simultaneously by the MRI technique. As mentioned above, this TEPA-based qubit formation is a unique technique that can be used for quantum computation, such as that shown in Fig. 2 for the two-qubit entanglement, but can easily be extended to multi-qubit quantum entanglement.

In short, MRI-based qubit generation can be noted as a System Hamiltonian and two external Control Hamiltonians[6,7]. First, the time evolution of the magnetization due to the System Hamiltonian with a static magnetic field $B_0$ can be written as:

$$|\Psi_s\rangle = e^{i H_0 t}|\Psi_0\rangle = e^{-i\frac{1}{2}\hbar\omega_0 t \sigma_z} = |\Psi_0\rangle$$

... Eq. 3

where $H_0 = -\frac{1}{2}\hbar\omega_0\sigma_z$ is a System Hamiltonian with an external static magnetic field of $B_0$. $\omega_0$ is the resonance frequency or Larmor frequency, which is given as $\omega_0 = \gamma B_0$, where $\gamma$ is the gyromagnetic ratio, $\hbar$ is Planck constant divided by $2\pi$, $\sigma_z$ is the z-directional Pauli matrix, and $|\Psi_0\rangle$ is the initial state.

**Gradient and Control Hamiltonians with a main RF coil and multiple individual Q-coils**

For the magnetization control for each qubit, two Control Hamiltonians are used, namely, the main control RF excitation for all qubits and multiple small Q-coil RF excitations for each qubit. First, the initial main Control Hamiltonian is given by:

$$H_c^m = -\frac{1}{2}\hbar\omega_{RF}\sigma_x \quad \text{... Eq. 4}$$

where $\omega_{RF}$ is the main control RF developed by the $B_1$ field ($\omega_{RF} = \gamma B_1$), with a bandwidth of $n\Delta\omega_0$ applied transversally to all qubits. The resulting time evolution due to the initial main Control Hamiltonian is:

$$(|\Psi_m\rangle)_{t=t_{0+}} = e^{i H_c^m \Delta t}|\Psi_s\rangle$$

$$= e^{-i\frac{1}{2}\hbar\omega_{RF}\Delta t \sigma_x}|\Psi_s\rangle \quad \text{... Eq. 5}$$

where $\sigma_x$ is the Pauli matrix in the x direction, as a rotating axis, and $\omega_{RF}\Delta t = 90°$ (or $180°$) depending on the situation, $\Delta t$ is the RF pulse length at Q-coils, and $i$ is the $\sqrt{-1}$.

Second, further individual Q-coils select the desired qubit or qubits and controls for the spin-echo signal generation. However, each Q-coil will have a different frequency $\omega_i$ and a series of $180°$ pulses, with the readout pulse determined by the desired probability amplitude values of each desired qubit. The second control Hamiltonian and the resulting time varying magnetization is given as:

$$H_c^Q = -\frac{1}{2}\hbar\omega_{RF}^j\sigma_x$$

... Eq. 6 ($j^{th}$-qubit Control Hamiltonian)

$$(|\Psi_m\rangle)_j = e^{i H_c^m \Delta t}|\Psi_m\rangle$$

$$= e^{-i\frac{1}{2}\hbar\omega_{RF}^j\Delta t \sigma_x}|\Psi_m\rangle$$

... Eq. 7 ($j^{th}$-qubit state due to individual Q-coil Control Hamiltonian)

where $|\Psi_m\rangle$ is the state developed after the first main control RF pulse ($90°$), $\sigma_x$ is the Pauli matrix in the x direction, as a rotating axis, $\Delta t$ is the pulse length of the RF pulse at the Q-coil for the formation of the $180°$ RF pulse, $\omega_{RF}^j$ is the center frequency of $j^{th}$ qubit, and $\omega_{RF}^j\Delta t = 180°$ for generation of the series of spin-echo signals.

**Qubit states, spin-echo series, and TEPA**

For a selected qubit, one of the spin-echo signals will be selected to produce a desired qubit state, the desired probability amplitudes, α and β, which are derived from one of the spin-echo signals, and which is read out at a selected time, i.e., a gate readout time $t = t_{i+}$ via a readout gate (RG1, RG2, etc.). This TEPA signal is the main tool for creating various qubits or qubit states, each of which has a unique set of values of probability amplitudes, which can be written as:

$$(|\Psi_m>)_{j, t=t_{i+}} = e^{-i H_c^Q \Delta t} |\Psi_m>$$

$$= e^{-i \frac{1}{2}\hbar \omega_{RF}^j \Delta t \sigma_x} |\Psi_m>$$

... Eq. 8 ($j^{th}$ *qubit spin-echo signal at* $t = t_{i+}$)

where $\sigma_x$ is the Pauli matrix in the x direction, as a rotating axis, $\omega_{RF}^j$ is the frequency of the $j^{th}$ qubit, and $\omega_{RF}^j = 180°$ is for spin-echo signal generation. Equation 8 is, therefore, one of the spin-echo signals representing the qubit state, the desired probability amplitude values, which is also the same as one of the spin-echo signals ($i^{th}$ spin-echo) that is read out at the selected time $t = t_{i+}$ via gating at the readout gate (RG1, RG1, etc.) of the respective qubit. In **Fig. 4**, one of the echo signals, the TEPA signal selected by the readout gate at time $t_{3+}$, is illustrated with a table that can either be calculated or experimentally obtained. Outputs of the qubits can then be applied to quantum computing together with quantum logic gates, such as the one shown in **Fig. 2**.

**DISCUSSION**

We have here described how to generate multiple qubits by using an MRI "gradient" together with a main control RF coil and pulse, and multiple of Q-coils for spin-echo generation, and subsequent readout of the spin-echo signals for the formation of a desired qubit state with probability amplitudes. This unique qubit state production technique with time encoding allowed us to generate multiple qubits of desired states or values, which we termed the TEPA technique, with which quantum computation can be performed. MR gradient-based qubit generation also allowed us to either increase or decrease the number of qubits, by simply extending the number of Q-coils with a suitable gradient[22]. Increasing the number of qubits would often be useful for error correction in quantum computing[7,8].

In addition to the two basic key techniques, gradient with multiple Q-coils and the TEPA technique with spin-echoes, the spin-echo technique further allowed us to recover the collapsed signal. This technique for recovery of lost qubit content is another unique technique of the MRI-based qubit generation approach. In addition, the MR-qubit technique will offer great flexibility in data processing, since the sample (H-Bar) can be placed at room temperature, allowing us to control and manipulate the qubits if necessary, unlike existing qubits that are either placed in a highly controlled, low temperature (ultra-low mK temperature) or high-vacuum environments, which are difficult to provide, costly, and time-consuming. In addition, the whole platform can be constructed on a microchip, which is easy to fabricate with existing semiconductor technology. It also should be mentioned that the input and output electronics are readily available in the current semiconductor technology.

In conclusion, we proposed a new MRI-based multiple qubit-generation technique with which one can readily develop a multi-qubit quantum computing platform. Its essential components are the following: First, the multi-qubit generation process, using "MRI gradient" incorporation with two control coils: one for the main control RF, such as the 90° RF broadband pulse for all the qubits, and the other the small, multiple control Q-coils for generation of individual qubits. After selection of individual qubits, spin-echo signals are generated for each qubit. Second, the development of the TEPA encoding for each qubit, with which one can control the probability amplitudes, α and β, of the qubit, the qubit states, as a function of time, and the spin-echo time or the readout gate opening time (such as $t_{1+}$ and $t_{3+}$). Third, the H-bar sample can be placed in a room temperature environment, unlike other superconducting qubits, which require ultra-low temperature (10–40 mK). It should also be mentioned that the many signal manipulating techniques developed in the field of MRI technology in the last several decades, such as multi-echo and stimulated echo techniques, could help further develop MR-qubit-based computational methods (e.g., error correction) by development of additional multi-qubit generation[19-25]. Since this method could easily generate multiples of qubits and allows us to encode by the spin-echo signal readout, both sequentially arranged and randomly organized qubits can easily be generated for computation and entanglement for communication.

**ACKNOWLEDGEMENTS**

The authors would like to thank to the colleagues and external reviewers who have provided essential information on quantum computation and quantum qubit developments, from superconducting qubits to ion-trap and quantum computing strategy. We thank Dr. Jaewan Kim, Korea Institute of Advanced Study; Dr. Eunmi Chae, Dept. of Physics, Korea University; Dr. Jun Heo, Dept. of Electrical and Electronics, Korea University; Dr. Yunwook Chung, Sungkyunkwan University; Dr. Yong Ho Lee, Korea Research Institute of Standard and Science; Dr. Sangwook Han, Korea Institute of Science and Technology; and Dr. Jaewook Ahn, Dept. of Physics, Korea Advanced Institute of Science and Technology.

This work was supported in part by the Brain Research Program of the National Research Foundation of Korea (NRF), which is funded by the Ministry of Science and ICT (2017MC7A1049026).

**AUTHOR CONTRIBUTIONS**

Z.H.C. contributed in the areas of MRI and MRI-based quantum qubit generation. Y.D.S. and H.J.J. contributed in the areas of MRI-related technical developments. Y.B.K. and S.H.P. contributed in the discussion related to quantum computing-related topics. D.H.S. made contributions with respect to the chemical aspects of NMR. H.G.L. made contributions in the field of superconducting magnets, particularly the possible use of the high temperature superconducting magnet for quantum computing platform.

Supplementary Table 1. Quantum qubit Generation Procedures or Steps :

Quantum qubit Generation Steps:

*After Application of Static or System Magnetic Field $B_0$*

1. Apply Gradient Field $G_z(z)$.
2. Apply the first Control RF ($90^0$) to all the qubits.
3. Select the desired qubits (QB1, 2, . . . ) by applying a second Control RF (180°) pulse series to desired or selected Q-Coils for spin-echo generation.
4. To generate the desired qubit state (α and β), select pulses at the Gate Readout time of the echo signal $t_{i+}$ ($t_{1+}$, $t_{2+}$, …) based on the conversion table of the probability amplitude*.
5. Spin echo signals are then converted to qubit states of the selected qubits and finally outputs are coupled to the Quantum Computing Logic Gates or Circuit for computation.

\* The conversion table is either calculated based on the T1 value or from experimentally measured data.